\documentclass[%
 aip,
 amsmath,amssymb,
 reprint,%
]{revtex4-1}

\usepackage{graphicx}
\usepackage{dcolumn}
\usepackage{bm}

\usepackage[utf8]{inputenc}
\usepackage[T1]{fontenc}
\usepackage{mathptmx}
\usepackage{etoolbox}
\usepackage{pifont}
\usepackage{upgreek}
\usepackage{float}
\usepackage[colorlinks=false, pdfborder={0 0 0}]{hyperref}
\makeatletter
\def\@email#1#2{%
 \endgroup
 \patchcmd{\titleblock@produce}
  {\frontmatter@RRAPformat}
  {\frontmatter@RRAPformat{\produce@RRAP{*#1\href{mailto:#2}{#2}}}\frontmatter@RRAPformat}
  {}{}
}%
\makeatother
\begin{document}

\preprint{AIP/123-QED}

\title{Quantum Storage of Frequency-Multiplexed Photons Exhibiting Nonclassical Correlations with Telecom C-Band Photons}
\author{Hiroki Tateishi}
\email{tateishi-hiroki-tn@ynu.jp}
\affiliation{Department of Physics, Yokohama National University, 79-5 Tokiwadai, Hodogaya-ku, Yokohama 240-8501,}

\author{Daisuke Yoshida}%
\affiliation{IMS, Yokohama National University, 79-5 Tokiwadai, Hodogaya-ku, Yokohama 240-8501,}
\affiliation{LQUOM, Inc.}

\author{Tomoki Tsuno}
\affiliation{Department of Physics, Yokohama National University, 79-5 Tokiwadai, Hodogaya-ku, Yokohama 240-8501,}
\affiliation{LQUOM, Inc.}

\author{Takuto Nihashi}%
\affiliation{Department of Physics, Yokohama National University, 79-5 Tokiwadai, Hodogaya-ku, Yokohama 240-8501,}

\author{Ryoma Komatsudaira}%
\affiliation{Department of Physics, Yokohama National University, 79-5 Tokiwadai, Hodogaya-ku, Yokohama 240-8501,}

\author{Daisuke Akamatsu}%
\affiliation{Department of Physics, Yokohama National University, 79-5 Tokiwadai, Hodogaya-ku, Yokohama 240-8501,}

\author{Feng-Lei Hong}%
\affiliation{Department of Physics, Yokohama National University, 79-5 Tokiwadai, Hodogaya-ku, Yokohama 240-8501,}
\affiliation{IMS, Yokohama National University, 79-5 Tokiwadai, Hodogaya-ku, Yokohama 240-8501,}

\author{Koji Nagano}%
\affiliation{IMS, Yokohama National University, 79-5 Tokiwadai, Hodogaya-ku, Yokohama 240-8501,}
\affiliation{LQUOM, Inc.}

\author{Tomoyuki Horikiri}%
\email{horikiri-tomoyuki-bh@ynu.ac.jp}
\affiliation{Department of Physics, Yokohama National University, 79-5 Tokiwadai, Hodogaya-ku, Yokohama 240-8501,}
\affiliation{IMS, Yokohama National University, 79-5 Tokiwadai, Hodogaya-ku, Yokohama 240-8501,}
\affiliation{LQUOM, Inc.}

\begin{abstract}
Multiplexing is essential for improving entanglement distribution rates in quantum communication. Frequency multiplexing provides a promising and scalable path toward large-capacity quantum networks. Further progress requires increasing the number of frequency modes and developing broadband photon-pair sources and quantum memories that are spectrally compatible. Here, we report the integration of a cavity-enhanced spontaneous parametric down-conversion source in the telecom C-band with a frequency-multiplexed atomic frequency comb memory. The bow-tie cavity source was simultaneously resonant at 606 nm and 1550 nm, generating non-degenerate photon pairs exhibiting a clustered frequency-comb spectrum. The atomic frequency comb memory, implemented in Praseodymium-doped Yttrium Orthosilicate crystals, provided up to 83 frequency modes with 123 MHz spacing and enabled broadband storage of 606 nm signal photons. By filtering the main cluster, we obtained $32.7 \pm 4.8$ effective modes, as confirmed from coincidence measurements. Importantly, we observed strong nonclassical correlations after storage, with cross-correlation values of $g_{s,i}^{(2)} = 8.1\pm0.7$. Our experimental results demonstrate the feasibility of integrating cavity-enhanced photon-pair sources with rare-earth-ion-doped solid-state memories. The integration reveals a high frequency multiplicity that is essential for scalable quantum networks.
\end{abstract}
\maketitle

Entanglement distribution between quantum memories has attracted significant attention as a fundamental technology for quantum internet\cite{Kimble08} with applications including distributed quantum computation\cite{Cirac99}, blind quantum computation\cite{broadbent09}, and world clocks with unprecedented stability and accuracy\cite{Komar14}. Multiplexing is crucial to improve the distribution rate, and extensive studies have been conducted toward improving the distribution rate\cite{Jing24,Sinclair14,Saglamyurek16,Yang18}. One promising scheme employs photon-pair sources (PPSs) and absorptive quantum memories\cite{Rivera21,Simon07}, where each node is equipped with both components. Among the generated photon pairs (signal and idler photons), the signal photon is absorbed and stored in the local memory, while the idler photon is transmitted to a central station. A successful Bell-state measurement at this station provides a heralding signal that establishes entanglement between the remote memories. This entanglement-distributuion approach has been demonstrated with temporal multiplexing\cite{Liu21,Rivera21,Hänni25}, where the atomic frequency comb (AFC) scheme has often been employed as the quantum memory and spontaneous parametric down-conversion (SPDC) has been used as the photon-pair source. AFC is a quantum memory scheme based on the creation of a periodic comb-like absorption structure within the inhomogeneous broadening of a rare-earth-ion-doped crystal using spectral hole-burning. In this scheme, an absorbed photon is mapped onto a collective atomic excitation and rephased after a time equal to the inverse of the comb spacing. As a result, the photon is retrieved as an echo. This periodic comb structure enables temporal multiplexing because the atomic excitations periodically rephase, allowing multiple temporal modes to be stored and retrieved.

To enhance the distribution rates, extension to the frequency domain is a next step. The AFC based on Praseodymium-doped Yttrium Orthosilicate (Pr:YSO) crystals provides a promising for frequency multiplexing. Pr:YSO is known to exhibit a narrow homogeneous broadening on the order of kHz and a broad inhomogeneous broadening of approximately 10 GHz\cite{Equall95}, which makes it well suited for multiplexing in frequency domains. In addition, as described above, the periodic comb structure of the AFC enables temporal multiplexing.
Theoretically, an AFC with a bandwidth of 4 MHz can support about 30 temporal modes and more than 100 frequency modes with $\sim100$ MHz spacing\cite{Ortu22}, indicating its strong potential as a multidimensional multiplexed memory.

For coupling to frequency-multiplexed memories, cavity-enhanced SPDC (cSPDC) is particularly attractive. In the cSPDC approach, a nonlinear crystal is placed inside an optical cavity, which enhances the spectral brightness while generating photon pairs in a frequency-comb structure defined by the longitudinal modes of the cavity\cite{Ou99}. When the cavity mode frequencies match the AFC frequency modes, efficient simultaneous storage of signal photons across multiple frequency modes can be achieved. The coupling of such a cSPDC source with a frequency-multiplexed AFC memory supporting up to 15 modes was demonstrated\cite{Seri19}, confirming the technical feasibility of the cSPDC–AFC integration. More recent studies have further advanced this integration\cite{Ito23}.

In this work, we present the first realization of frequency-multiplexed coupling between a cavity-enhanced PPS that generates non-degenerate photons at 606 nm and 1550 nm (telecom C-band), and an AFC quantum memory. Extending the approach to the telecom C-band is particularly important because photons at the telecom C-band experience minimal transmission loss in optical fibers, and they are therefore well suited for large-scale quantum networks. To date, demonstrations of cSPDC–memory coupling have been limited to the S- and E-bands\cite{Ito23,Rivera21,Hänni25,Seri19}.

\begin{figure}[H]
\includegraphics[width=0.5\textwidth]{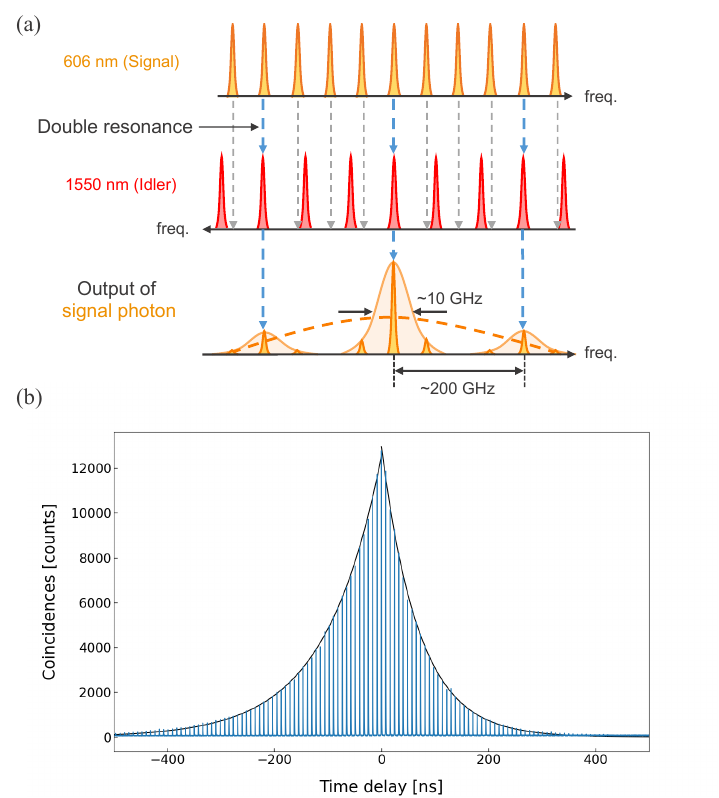}
\caption{\label{fig1}(a) Cluster structure of our PPS. Since the SPDC output bandwidth is restricted to the frequencies where the two wavelengths are resonant simultaneously (double resonance), the output spectrum of each photon forms a cluster structure. The central cluster (main cluster) has a width of approximately 10 GHz. The orange dashed line indicates the broadband spectral envelope of SPDC without the cavity, illustrating the full phase-matching bandwidth before spectral narrowing by the cavity. (b) Time correlation of photon pairs generated by the PPS without cavity-length locking. The histogram shows the time difference between the detection of an idler photon and that of a signal photon, and the black line represents a fit to the envelope. The pump power was 0.1 mW, and the measurement time was 10 min.}
\end{figure}

In the following, we first describe the frequency-multiplexed PPS developed in this work. 
The PPS generated photon pairs at 606 nm (signal) and 1550 nm (idler) through non-degenerate SPDC using a periodically poled potassium titanyl phosphate (PPKTP) crystal under type-0 quasi-phase matching\cite{Aizawa23}. 
In this study, the crystal was placed inside a bow-tie cavity to restrict the photon emission spectrum to the cavity resonant frequencies, thereby enhancing the spectral brightness through cSPDC. As the SPDC pump, we used light from an external cavity diode laser (ECDL) at 871 nm, which was converted to 436 nm through second-harmonic generation (SHG). Because non-degenerate SPDC was employed, a difference in free spectral ranges (FSRs) arose between the two output wavelengths due to the difference in the optical path-length caused by wavelength dispersion in the crystal. As a result, the two-photon spectrum exhibited a cluster structure\cite{Luo15}, as shown in Fig.~\ref{fig1}(a). The cluster width and spacing were estimated to be approximately 10 GHz and 200 GHz, respectively, based on the model reported in Ref.~\cite{Rieländer16}.

Figure~\ref{fig1}(b) shows the time correlation of the photon pairs generated from the PPS used in our experiment. For cSPDC, the second-order cross-correlation function $G_{s,i}^{(2)}(\tau)$ between the signal and idler photons in cSPDC can be expressed as follows\cite{Luo15}:
\begin{align}
G^{(2)}_{s,i}(\tau) \propto & \; u(\tau) e^{-2\pi\Delta\nu_s\tau}
\left[
\sum_{n=1}^N s_n^2
+ \sum_{n=1}^{N-1} \sum_{j=1}^{N-n}
2 s_j s_{j+n} \cos\!\left(n\Delta_{F_s}\tau\right)
\right] \nonumber \\
& + u(-\tau) e^{2\pi\Delta\nu_i\tau}
\left[
\sum_{n=1}^N s_n^2
+ \sum_{n=1}^{N-1} \sum_{j=1}^{N-n}
2 s_j s_{j+n} \cos\!\left(n\Delta_{F_i}\tau\right)
\right],
\end{align}
where $u(\tau)$ is the step function, $\Delta\nu_s$ and $\Delta\nu_i$ are the linewidths of the cavity at the signal and idler frequencies, $\Delta_{F_{s(i)}}=2\pi\mathrm{FSR}_{s(i)}$, $N$ is the number of frequency modes, and $s_n$ are different phase-matching weights. In particular, the exponential factor determines the envelope of the coincidence counts, as shown by the black solid line in Fig.~\ref{fig1}(b). By fitting the experimental data with an exponential function, the linewidth of the signal and idler photons, $\Delta\nu_s$ and $\Delta\nu_i$, were evaluated to be 2.28 MHz and 1.52 MHz, respectively. Furthermore, the comb-like peaks in Fig.~\ref{fig1}(b) correspond to the difference in the number of cavity round trips between the signal and idler photons. Because different numbers of round trips lead to discrete changes in time delay, peaks appear periodically with respect to time delay $\Delta\tau$. Since the peak interval corresponds to the cavity round-trip time, we determined the temporal spacing between adjacent peaks by averaging the intervals from the 0~ns (0th) peak to the 30th peak in the positive delay direction. As a result, the mean interval was found to be $\Delta \,t = 8.13 \pm 0.01~\mathrm{ns}$, and the cavity FSR was estimated as $\mathrm{FSR} =1⁄\Delta \,t=123.0\pm 0.2\,\mathrm{MHz}$.

To stabilize the cavity at the signal wavelength, a reference light at 606 nm was injected into the PPS in addition to the pump light. Without active stabilization, small drifts in the cavity length would detune the cavity resonance from the AFC bandwidth. This detuning would hinder efficient photon storage. The reference light was also used for difference-frequency generation (DFG) with the 436 nm pump.
As a result, 1550 nm light was then generated and simultaneously brought into resonance with the reference light. This simultaneous resonance of the two wavelengths is referred to as double resonance \cite{Luo15}. The crystal temperature was tuned to maximize the DFG intensity, thereby optimizing the phase-matching condition. This tuning effectively adjusted the signal wavelength (606 nm) so that its central frequency matched that of the reference light.

Next, we describe the frequency-multiplexed memory developed in this study. 
We employed a Pr:YSO crystal with a doping concentration of 0.05\%, which was placed inside a cryostat cooled to approximately 3 K. We performed spectral hole burning using a preparation laser externally modulated by an acousto-optic modulator (AOM), forming an AFC with a comb spacing of 920 kHz, which corresponds to a storage time of approximately $1.1~\mathrm{\upmu s}$. 
To achieve multimode operation, we applied a multi-tone signal to a single electro-optic modulator to generate sidebands by phase modulation. 
Using this method, the frequency distribution of the laser for AFC preparation was shaped into a comb with 123 MHz spacing, enabling the formation of up to 83 AFC modes across a bandwidth of about 10 GHz. 
The 123 MHz spacing was matched to the longitudinal mode spacing of the PPS, as shown in Fig.~\ref{fig2} (see Supplementary Material A for the details of the preparation of the frequency-multiplexed AFC).
For the AFC preparation laser, we used the SHG output of an ECDL at 1212 nm. 
The 1212 nm laser was frequency-stabilized to an optical frequency comb\cite{Asahina19}.
This optical frequency comb was in turn stabilized to an Nd:YAG laser that had been locked to the absorption line of iodine molecules, serving as an external reference. The iodine-stabilized Nd:YAG laser provided a highly stable frequency reference, which ensured long-term frequency stability of the optical frequency comb.

The spectrum of photon pairs generated by cSPDC is determined by the resonant spectrum of the cavity, and the 606 nm laser used for AFC preparation was employed as the reference light for cavity length stabilization. 
This ensured precise frequency matching between the absorption wavelength that defines the AFC and the wavelength of the photons incident on the AFC. 

\begin{figure}[H]
\includegraphics[width=0.45\textwidth]{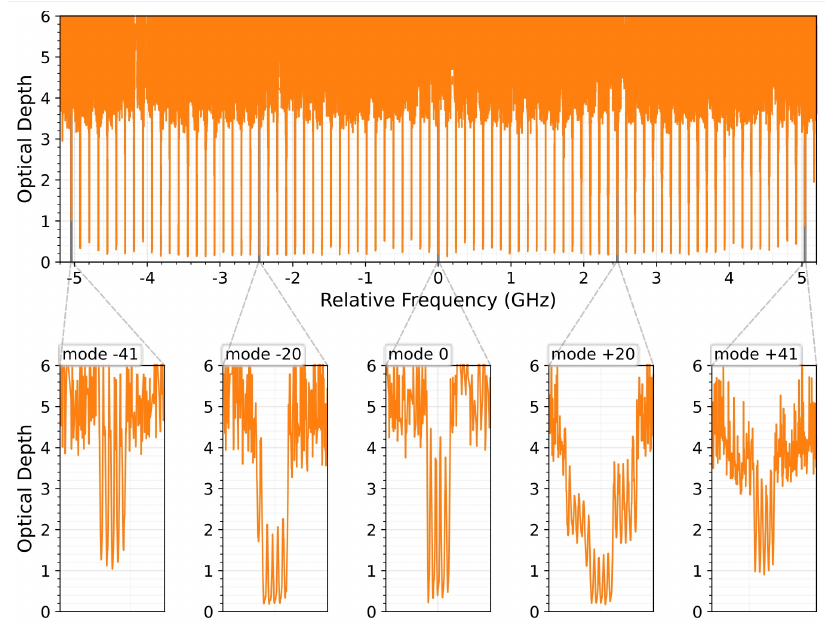}
\caption{\label{fig2}Absorption spectrum of the Pr:YSO crystal showing an AFC with approximately 83 frequency modes.
Insets show 20-MHz-wide magnified views of representative modes $0$, $\pm20$, and $\pm41$.}
\end{figure}

\begin{figure*}
\includegraphics[width=\textwidth]{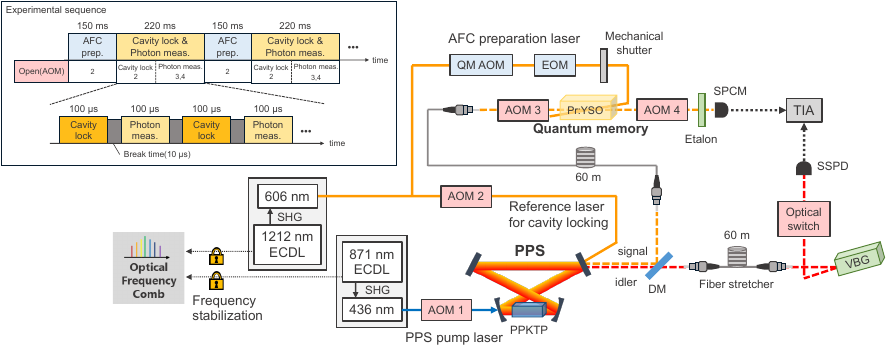}
\caption{\label{fig3}Schematic of the experimental setup. The PPS pump laser (436 nm) is incident on a PPKTP crystal inside the cavity. The signal and idler photons are separated by a DM and directed to their respective paths. Among them, the 1550 nm idler photons are reflected by a VBG so that only the main cluster is selectively filtered, and they are detected by an SSPD. The 606 nm signal photons are stored in the quantum memory, and the retrieved photons are subsequently detected by an SPCM. An etalon is placed in front of the SPCM to reduce noise by filtering out photons outside the inhomogeneous broadening of Pr:YSO. Both lasers used in this experiment are frequency-stabilized using an optical frequency comb. The inset shows the experimental sequence. This sequence is cyclically repeated and indicates the timing when the AOM shutters in the setup are opened.}
\end{figure*}

We now describe the coupling experiment between the developed frequency-multiplexed PPS and quantum memory, the main subject of this study. A schematic of the experimental setup is shown in Fig.~\ref{fig3}. Photon pairs generated from the PPS are separated by a dichroic mirror (DM), and the signal photons are stored in the quantum memory. The retrieved photons pass through an etalon (ETASQ1-12.5C0.6-1-606-R90, SIGMAKOKI) with a transmission bandwidth of 5.58 GHz and an FSR of 166 GHz (transmittance:50\%), and are subsequently detected by a single-photon counting module (SPCM, SPDMH2F, THORLABS). The idler photons are filtered by diffraction using a volume Bragg grating (VBG, BP-1550, OptiGrade) with a bandwidth of $\sim10$ GHz (reflectivity:92\%). These frequency filters selectively extract the photon pairs belonging to the main cluster. The idler photons are subsequently detected by a superconducting single-photon detector (SSPD, ECOPRS-CCR-TW85, SCONTEL). Finally, the time difference between the detections of the signal and idler photons is measured by a time interval analyzer (TIA, ID1000-MASTER, ID Quantique).

In Fig.~\ref{fig3}, AOMs (1–4) and the optical switch (CrystaLatch, Agiltron) serve as intensity modulators and temporal gates. 
AOM~1 controls the intensity of the pump light used for generating the DFG light and for photon measurements. 
AOM~2 prevents the reference laser for cavity-length locking from entering the SPCM during photon measurements.
AOM~3 is gated by a signal from the SSPD when idler photons are detected. As a result, spurious counts during photon echo measurements are suppressed, and the noise level is reduced (see Supplementary Material B for details of the reduction scheme). 
AOM~4 and the optical switch are placed before the detectors. They block classical light (reference laser and DFG light) that could otherwise enter the detectors during cavity-length locking. We used the AOMs and the optical switch to alternate between cavity-length locking and photon measurements, instead of employing a mechanical chopper\cite{Rivera21}, in order to suppress mechanical vibrations.

The inset in Fig.~\ref{fig3} shows the overall experimental sequence, which consists of two main stages: the AFC preparation stage and the memory operation stage (photon storage and retrieval). 
In the AFC preparation stage, an AFC is created in the Pr:YSO crystal as described above. The subsequent memory operation stage is further divided into two sub-stages, photon measurement and cavity locking, which alternate in a $100~\mathrm{\upmu s}$ cycle. 
A break time of $10~\mathrm{\upmu s}$ is inserted between the two sub-stages to suppress noise caused by residual reference light used from the cavity-locking. 
The inset also shows the timing when the AOM shutters (1–4) are opened during the AFC preparation and memory operation stages.

\begin{figure*}
\includegraphics[width=\textwidth]{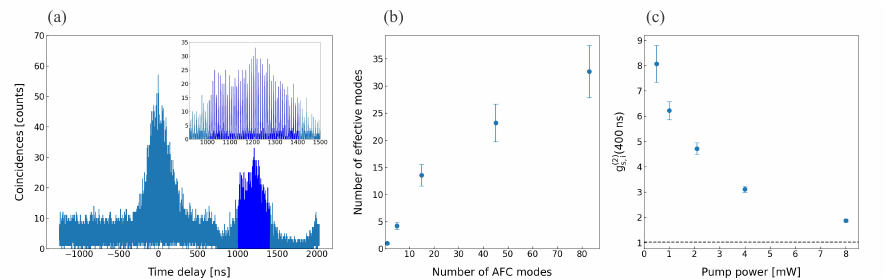}
\caption{\label{fig4}(a) Time correlation between idler photons and signal photons retrieved from the AFC. The inset shows the comb structure. The SPDC pump power was 1 mW, and the measurement time was 2 h. (b) Number of effective modes as a function of the number of AFC modes created in the Pr:YSO crystal. The SPDC pump power was 2.2 mW. (c) Second-order cross-correlation function $g_{s,i}^{(2)}(400~\mathrm{ns} )$ as a function of the pump power in cSPDC. The black dashed line indicates the classical limit ($g_{s,i}^{(2)} = 1.03$) for 33 modes.}
\end{figure*}

Figure~\ref{fig4}(a) shows the time correlation between the idler photons and the photons retrieved from the AFC. 
In this experiment, an AFC with 83 frequency modes was created in the Pr:YSO crystal.
The time bin size of the TIA was 0.2~ns, the pump power was 1 mW, and the measurement time was 2 h. 
The comb structure observed around 0~ns corresponds to coincidences between idler photons and signal photons that were directly transmitted without being absorbed by the AFC. The step-like features in the noise floor around 700~ns and 1900~ns correspond to the activation and deactivation of the noise reduction system implemented with AOM~3 in Fig.~\ref{fig3}.
The comb structure observed around 1200~ns corresponds to coincidences between the echo photons and the idler photons, as the storage time of the AFC in this experiment is approximately $1.1~\mathrm{\upmu s}$. Considering the range in which the comb structure of the echo photons can be clearly distinguished, we set the coincidence time window to $\Delta \tau = 400~\mathrm{ns}$. The shaded region in Fig.~\ref{fig4}(a) corresponds to this time window.

Figure~\ref{fig4}(b) shows the number of effective modes of photons stored in the AFC when the number of modes of the AFC created in the Pr:YSO crystal was varied (see Supplementary Fig.~S3 for profiles of several frequency-multiplexed AFCs).
The number of effective modes was defined by dividing the obtained rate by that measured when the AFC was prepared with a single mode. For each measurement, the coincidence rate was obtained by subtracting the noise floor from the coincidence counts within the 400~ns time window and dividing by the measurement time. 
Here, the noise floor was taken as the average value in the 1600–1800~ns region, which is sufficiently separated from both the decay tail of the comb structure and the turn-on and turn-off periods of the noise-reduction system. In this experiment, when an AFC with up to 83 modes was created, the number of effective modes was $32.7 \pm 4.8$.

As a performance evaluation of the PPS and the AFC, we use the normalized second-order cross-correlation function, defined as $g_{s,i}^{(2)}(\Delta \tau) = \frac{p_{s,i}}{p_s p_i}$, where $p_{s,i}$ is the coincidence probability between signal and idler photons, and $p_{s,(i)}$ is the probability of detecting signal (idler) photons individually within the time window $\Delta\tau$. 
In Fig.~\ref{fig4}(a), we obtained $g_{s,i}^{(2)} = 6.2\pm0.4$. Fig.~\ref{fig4}(c) shows $g_{s,i}^{(2)}$ as a function of the pump power in cSPDC. For a given number of modes $N$, the classical limit of the cross-correlation function is expressed as follows\cite{Christ11}:
\begin{equation}
g^{(2)}_{s,i} \leq 1 + \frac{1}{N}.
\label{eq2}
\end{equation}
As mentioned above, the number of effective modes in this study was estimated to be 33, which yields a classical limit of 1.03, as shown by the black dashed line in Fig.~\ref{fig4}(c). Therefore, the measured values exceed this limit for all pump powers, representing nonclassical behavior.
In particular, at a pump power of 0.5 mW, a maximum correlation of $g_{s,i}^{(2)} = 8.1\pm0.7$ was observed. Furthermore, the $g_{s,i}^{(2)}$ values obtained after storage were almost identical to those measured without storage, and no significant degradation was observed due to the storage process.

In this study, even though AFCs with up to 83 modes (equivalent to 9.96 GHz in frequency) were created, the number of effective modes was limited to 33 (4.06 GHz). The number of effective modes did not increase linearly with the number of AFC modes. 
The deviation from linearity can be attributed to the transmission bandwidth of the etalon filter (approximately 5.6 GHz), which restricted the number of transmitted modes when the AFC supported 83 modes (9.96 GHz). A smaller deviation from linearity was also observed when the AFC bandwidth was 45 modes (5.5 GHz), which can be explained by the non-uniform spectral distribution of the photon pairs and the non-uniform storage efficiency of the AFC. If one aims to increase the number of effective modes, several improvements can be considered. One possible approach is to optimize the filtering bandwidth for frequency multiplexing to increase the number of transmitted signal photons. 
In addition, the spectrum of the AFC preparation light can be further flattened to maintain uniform storage efficiency even for modes that are detuned from the central frequency. Improvements aimed at increasing the effective multiplicity are left for future work, with the goal of achieving a higher gain in frequency multiplexing.

We conclude that our work represents an important step toward realizing entanglement generation through the integration of a frequency-multiplexed PPS and a quantum memory, marking the first demonstration in the telecom C-band. The generated photons were successfully stored in a quantum memory with an AFC structure of up to 83 modes.
Through frequency multiplexing, a gain of up to 33 in the coincidence rate was achieved compared with single-mode operation. Furthermore we confirmed that a high correlation of $g_{s,i}^{(2)} = 8.1 \pm 0.7$ was maintained even after AFC storage within a 400-ns time window.

\vspace{1\baselineskip}
We acknowledge Ryosuke Hagiwara and Yuta Isawa for their technical support during the experiments, and Akira Ozawa for valuable discussions. This work was supported by Ministry of Internal Affairs and Communications R\&D of ICT Priority Technology Project (JPMI00316), New Energy and Industrial Technology Development Organization Deep-Tech Startups Support Program, and JST Moonshot R\&D (JPMJMS226C).
T. Tsuno, D. Yoshida, K. Nagano and T. Horikiri have financial interests in LQUOM Inc.
\vspace{-\baselineskip}
\section*{Data Availability Statement}
The data that support the findings of this study are available from the corresponding author upon reasonable request.
\vspace{-\baselineskip}
\section*{References}
\nocite{*}
\bibliography{refs}
\end{document}